\documentclass[12pt]{article}
\usepackage{fleqn}

\newcommand{\AmS}{{\protect\the\textfont2
  A\kern-.1667em\lower.5ex\hbox{M}\kern-.125emS}}

\begin{document}

\section*{Inelastic Scattering of Tritium-Source Antineutrinos on Electrons of Germanium Atoms}

V. I. Kopeikin, L. A. Mikaelyan, and V. V. Sinev

{\it Russian Research Centre Kurchatov Institute,  Moscow, Russia}
       
\vspace{0.5cm}
Processes of the inelastic magnetic and weak scattering of tritium-beta-source antineutrinos 
on the bound electrons of a germanium atom are considered. The results obtained by calculating 
the spectra and cross sections are presented for the energy-transfer range between 1 eV and 
18 keV.
\vspace{1pc}


\section*{Introduction}
This study was performed in connection with the experiment planned by researchers from three 
institutions (Institute of Theoretical and Experimental Physics, Joint Institute for Nuclear 
Research, and the All-Russia Research Institute for Experimental Physics) to look for the magnetic 
moment of the electron neutrino [1]. This experiment is presumed to study the spectra of energy 
deposited in a detector material upon the inelastic scattering of tritium-source antineutrinos 
on atomic electrons. The sensitivity to the magnitude of the neutrino magnetic moment is 
assumed to be ${\mu} \sim 3\times 10^{-12}{\mu}_{B}$ (${\mu}_{B}$ is the Bohr magneton), which is nearly two orders of 
magnitude better than the upper limit, ${\mu} \le 2\times 10^{-10}{\mu}_{B}$, set in experiments performed at the 
Savannah River nuclear reactors, the Rovensk nuclear power plant, and Krasnoyarsk neutrino 
laboratory [2].

The discovery of the neutrino magnetic moment would suggest the existence of phenomena beyond 
the fundamentals of electroweak-interaction theory and would have far-reaching consequences 
for particles physics and astrophysics. (For a discussion on the issues concerning the 
motivation of searches for the neutrino magnetic moment in laboratory experiments and the 
status and prospects of investigations along these lines, see, for example, the original 
study of Neganov {\it et al}. [1], the review articles quoted in [3], and references therein.)

The magnetic moment can be sought by analyzing the measured spectrum of events of $\bar{{\nu}_e}e$ 
scattering. In the case of scattering on a free electron, the cross section for weak $\bar{{\nu}_e}e$ 
scattering tends to a finite limit, while the cross section for magnetic scattering grows 
indefinitely. Therefore, the experimental sensitivity to the magnitude of the neutrino 
magnetic moment increases with decreasing energy of measured events. As the energy of an 
event decreases, however, the effects of electron binding in target atoms begin to affect 
relevant cross sections and their energy dependence.

A feature peculiar to the experiment proposed in [ 1 ] is that, in the majority of cases, 
the energy $q$ deposited in the target upon the scattering of tritium neutrinos $\bar{{\nu}_e}$ is below 
1 keV. The experiment will employ germanium and silicon detectors that, owing to the 
internal-signal-amplification mechanism, will make it possible to perform measurements 
in the region of energies as low as those indicated immediately above.

In this study, we consider the inelastic scattering of tritium-source antineutrinos on 
germanium-atom electrons. The ensuing exposition is organized as follows. First, we recall 
the expression for the cross sections for elastic $\bar{{\nu}_e}e$. scattering, whereupon we briefly 
dwell on features of inelastic scattering and on the procedure for calculating relevant 
cross sections. Further, we display the results obtained by calculating the spectra of 
inelastic scattering on electrons of germanium-atom subshells and, finally, present the 
total spectra and integrated cross sections in a form that is suitable for a direct 
comparison with experimental data.

\section{Elastic and inelastic scattering of antineutrinos on  an electron}

1. The differential cross section for the magnetic scattering of a neutrino on a free electron at rest is proportional to the squared 
magnetic moment ${\mu}^2$ [4];
that is,
\begin{equation}
(d{\sigma}^{M}/dq)_{free}={\pi}r^{2}_{0}({\mu}/{\mu}_{B})^{2}(1/q-1/E),
\end{equation}
where ${\pi}r^{2}_{0} = 2.495\times 10^{-25}$ cm$^2$; $E$ is the incident-neutrino energy; and $q$ is the energy transfer, 
which coincides, in this case, with the recoil-electron kinetic energy.

The differential cross section for the electroweak scattering of an antineutrino on a free electron at rest has the form (see, for 
example, [5])

\begin{eqnarray}
(d{\sigma}^{W}/dq)_{free} =G^{2}_{F}(m/2{\pi})[4x^{4}+  \nonumber \\
+ (1+2x^{2})^{2}(1-q/E)^{2}-2x^{2}(1+2x^{2})mq/E^{2}] ,
\end{eqnarray}
where $m$ is the electron mass; $G^{2}_{F}(m/2{\pi}) = 4.31\times 10^{-48}$ cm$^2$keV; and 
$x^{2} = \sin^{2}{\theta}_{W} = 0.232$, ${\theta}_{W}$ being the Weinberg angle.

For a given value of primary neutrino energy $E$, the energy transfer to the electron involved is restricted by the kinematical 
condition
\begin{equation}
q \le q_{max} = 2E^{2}/(2E+m). 
\end{equation}

In the region $q > q_{max}$, free-scattering cross sections are zero. In the case of tritium-source antineutrinos $\bar{{\nu}_{e}}$ 
whose endpoint energy is 18.6 keV, $q_{jmax}$ = 1.26 keV.

With decreasing energy, the cross section for free magnetic scattering increases in proportion to $1/q$, while the cross section 
(2) tends to a finite limit of $1.016\times10^{-47}$ cm$^2$keV. At the magnetic-moment value of ${\mu} = 3\times 10^{-12}$, 
the spectra of weak and magnetic antineutrino ($\bar{{\nu}_{e}}$) scattering intersect at an energy transfer of $q \approx$ 0.3 keV.

2. In calculating the differential cross sections $d{\sigma}^{M}_{i}/dq$ and $d{\sigma}^{W}_{i}/dq$ for inelastic scattering 
on the electrons of the $i$-th shell, we use the approach developed in [6]. In the initial state, there are a germanium atom 
(Z = 32) and incident antineutrinos of energy $E$; in the final state, there are a knock-on electron, an outgoing neutrino, and 
a vacancy in one of the atomic shells. 
The wave functions for the target atom and the electron binding energies are calculated in the self-consistent relativistic 
Hartree-Fock-Dirac model. The wave functions for the knock-on electron can be found by numerically integrating the
Dirac equation in the same self-consistent potential. The energy-transfer spectra $S^{M}_{i}(q)$ and $S^{W}_{i}(q)$ are 
obtained as convolutions of the differential cross sections $d{\sigma}^{M}_{i}/dq$

\begin{table*}[htb]
\caption{Calculated electron binding energies ${\epsilon}_{i}$ (in keV) in germanium atomic shells ($n_i$, is the number of 
electrons in a shell)}
\label{table:1}
{\footnotesize
\begin{tabular}{c|c|c|c|c|c|c|c|c|c|c|c|}
\hline
 & $K$ & $L_{I}$ & $L_{II}$ & $L_{III}$ & $M_{I}$ & $M_{II}$ & $M_{III}$ & $M_{IV}$ & $M_{V}$ & $N_{I}$ & $N_{II}$ \\
 & $(1s_{1/2}$ & $(2s_{1/2}$ & $(2p_{1/2}$ & $(2p_{3/2}$ & $(3s_{1/2}$ & $(3p_{1/2}$ & $(3s_{3/2}$ & $(3d_{2/2}$ & $(3d_{5/2}$ & $(4s_{1/2}$ & $(4p_{1/2}$ \\
\hline
$n_{i}$ & 2 & 2 & 2 & 4 & 2 & 2 & 4 & 4 & 6 & 2 & 2 \\
${\epsilon}_{i}$ & 10.9 & 1.37 & 1.22 & 1.19 & 0.17 & 0.12 & 0.115 & 0.031 & 0.03 & (0.013) & (0.005) \\
\hline
\end{tabular}\\[2pt]
}
\end{table*}

and $d{\sigma}^{W}_{i}/dq$ with the spectrum ${\rho}(E)$ of the tritium-source antineutrinos (Fig. 1) that was calculated 
by using the Coulomb function [7]. The resulting spectra $S^{M,W}(q)$ for scattering on a germanium atom can be derived 
by taking a sum over the shells of the spectra $S^{M,W}_{i}$ with allowance for the number $n_i$, of electrons in each shell:
\begin{equation}
S^{M,W}(q) = \sum (n_{i}/Z)S^{M,W}_{i}(q).
\end{equation}

All the spectra $S^{M,W}_{i}$ and the resulting spectra $S^{M,W}(q)$ are normalized to one electron and are calculated in 
units of $10^{-47}$ cm$^2$/(keV electron). In all the calculations of cross sections and magnetic-scattering spectra, the 
neutrino magnetic moment is taken to be $5\times 10^{-12}{\mu}_{B}$.
The energy transfer $q$ to the atom involved in an inelastic-scattering event is the sum of the kinetic energy $T$ of the 
knock-on electron and the electron binding energy ${\epsilon}_i$, in the $i$th atomic shell:
\begin{equation}
q = T+{\epsilon}_i.
\end{equation}

The radiation generated upon the filling of the vacancy and the knock-on electron are absorbed in a detector material. As a 
result, the value determined experimentally for the energy of an event coincides with the energy transfer $q$ in this collision event.

The cross sections for weak and magnetic inelastic scattering on an electron of the $i$th shell vanish for $q \le {\epsilon}_i$. 
(We disregard transitions of the knock-on electron to discrete excited optical levels of the target atom.) In contrast to what is 
known for free scattering, these cross sections do not vanish at $q = q_{max}$ [see (3)]; in the "forbidden" region 
$q > q_{max}$, the inelastic-scattering cross sections decrease and vanish only at the $q$ value equal to the 
incident-antineutrino energy $E$.

The calculated electron binding energies in the shells of a free germanium atom (Table 1) agree with experimental data to 
within 5\%. This is not true only for four valence electrons, but their binding energy is so low that, in calculating cross sections 
and spectra, we use, for them, the formulas for scattering on free electrons.
 
\section{Discussion of the results for spectra and cross sections}

The scattering-event spectra $S^{M,W}_{i}(q)$ depends substantially on the binding energy ${\epsilon}_i$, in a shell, their 
shapes being different in the regions $q > q_{max}$ and $q < q_{max}$, where $q_{max} = 1-26$ keV is the kinematical 
limit given by (3).

\begin{table*}[htb]
\caption{Integrated cross sections $I^{M}(q)$ and $I^{W}(q)$ (in units of $10^{-47}$ cm$^2$/electron) for the inelastic 
magnetic (${\mu} = 5\times 10^{-12}$) and weak scattering of tritium neutrinos on germanium-atom electrons 
$\bar{{\nu}_{e}}e$ in the energy range 0.01-$q$ keV}
\label{table:2}
\begin{tabular}{c|c|c}
\hline
$q$, keV & $I^{M}(q)$ & $I^{W}(q)$ \\
\hline
0.02 & 0.054 & 0.0012 \\
0.03 & 0.0855 & 0.0025 \\
0.05 & 0.1372 & 0.0061 \\
0.1 & 0.2386 & 0.0173 \\
0.2 & 0.4046 & 0.0455 \\
0.35 & 0.5733 & 0.0892 \\
0.5 & 0.6771 & 0.1254 \\
0.75 & 0.7225 & 0.1671 \\
1. & 0.8184	 & 0.1907 \\
1.26 & 0.842 & 0.2031 \\
2.03 & 0.8771 & 0.2186 \\
3. & 0.887 & 0.2228 \\
5. & 0.8903	 & 0.2241 \\
10.94	 & 0.8908 & 0.2243 \\
18.6 & 0.8909 & 0.2243 \\
\hline
\end{tabular}\\[2pt]
\end{table*}

Figures 2 and 3 display the results obtained by calculating the spectra of inelastic scattering on electrons of the $L$ and $M$ 
germanium shells in the energy-transfer range $q$ = 1$-$4 keV, whose larger part belongs to the region $q > q_{max}$, which 
is forbidden for free scattering. With the exception of a small local enhancement near the threshold for the scattering of 
antineutrinos $\bar{{nu}_{e}}e$ on electrons of the $L$ shells, we observe a decrease in spectra with increasing energy $q$.
A general trend is that the lower the electron binding energy in a shell, the steeper this decrease.

In the energy region $q < q_{max}$ (Figs. 4, 5), $L$-shell electrons make virtually no contribution. This region is dominated 
by the scattering on electrons of $M$ atomic shells and on valence electrons. The spectra of magnetic scattering on $M$ shells 
in the immediate vicinity of the threshold lie much lower than the spectrum of scattering on a free electron. With increasing 
energy, the distinction between these spectra decreases fast; in a certain energy interval, they virtually coincide, while, for 
$q > 0.8$ keV, the inelastic-scattering spectra lie above the free-scattering spectrum. A similar behavior is also observed for 
weak scattering, but the spectra of elastic and inelastic scattering approach each other more slowly in this case.

The spectra $S^{M}(q)$ and $S^{W}(q)$ of magnetic and weak inelastic scattering that include the contributions from all 
germanium atomic shells are represented by the solid curves in Figs. 6 and 7. They can be used to perform a comparison 
with experimental data. The numbers of scattering events occurring within a given energy interval are determined by the integrals 
$I^{M}(q)$ and $I^{W}(q)$ of the spectra $S^{M}(q)$ and $S^{W}(q)$. Figure 8 and Table 2 display the integrals from 
the lower detection threshold of 0.01 keV to a variable upper boundary $q$.

>From the data given in Figs. 7 and 8 and in Table 2, it can be seen that, for the chosen value of the neutrino magnetic moment, 
the magnetic scattering is more intense than weak scattering over the entire energy region. It should be noted that nearly 
80\% of all magnetic-scattering events fall within the energy-transfer range from 0.05 to 1.26 keV; the energy region of 
$q > q_{max}$ involves only about 6\% of such scattering events; and their fraction in the interval from 0.01 to 0.05 keV 
amounts to about 15\%.

It is interesting to compare the "exact" spectra $S^{M,W}(q)$ and integrated cross sections $I^{M,W}(q)$ with the values 
obtained in the step-function approximation. In this approximation, it is assumed that the spectrum of inelastic scattering on an 
electron of the $i$th shell is coincident with the spectrum of scattering on a free electron for energies of $q > {\epsilon}_{i}$, 
and that it vanishes for $q < {\epsilon}_{i}$,. In this approximation, the "exact" expression (4) for the resulting spectra reduces 
to the form
\begin{equation}
S^{M,W}_{SF}(q) = S^{M,W}_{free}(q)  \sum (n_{i}/Z){\theta}(q-{\epsilon}_{i}).
\end{equation}
where summation is performed over all atomic shells and ${\theta}(q-{\epsilon}_{i})$ is the Heaviside step function equal to 
unity for $q > {\epsilon}_{i}$, and to zero for $q < {\epsilon}_{i}$. It was shown in [8] that, in the case where antineutrinos 
generated by a reactor and by a $^{90}$Sr-$^{90}$Y source are scattered on electrons of iodine and germanium atoms, the 
approximate spectra (6) agree with the spectra in (4) to within 2\% for the energy transfer $q$ ranging from 1-1.5 to 2 -3  keV.

>From the data shown in Fig. 6, it can be seen that the approximation specified by Eq. (6) is inadequate for both magnetic and 
weak scattering in the case of a soft spectrum of tritium antineutrinos. The integrated cross sections found in the step-function 
approximation exceed the corresponding "exact" values by about 25\%(Fig. 8).

\section*{ACKNOWLEDGMENTS}

We would like to emphasize the leading role of S.A. Fayans, who prematurely passed away in 2001, in developing and 
applying the procedure used in this study. We are grateful to L.N. Bogdanova and A.S. Starostin for stimulating discussions.
This work was supported by the Russian Foundation for Basic Research; by the program Leading Scientific Schools 
(grant No. 00-15-96708); and, in part, by the Department of Atomic Science and Technology of the Ministry for Atomic 
Industry of the Russian Federation (contract No. 6.25.19.19.02.969).

\section*{REFERENCES}
1. B. S. Neganove {\it et al.},Yad. Fiz. 64, 2033 (2001) [Phys. At.Nucl.64, 1948(2001)]. \\
2. F. Reines, H. Gurr, and H. Sobel, Phys. Rev. Lett. 37, 315 (1976); G. S. Vidyakin, V. N. Vyrodov, I. I. Gurevich {\it et al.}, 
Pis'ma Zh. Eksp. Teor. Fiz. 55, 212 (1992) [JETP Lett. 55,206 (1992)]; A. V. Derbin et al., Pis'ma Zh. Eksp. Teor. Fiz. 
57,755 (1993) [JETP Lett. 57,768 (1993)]. \\
3. A. V. Derbin, Yad. Fiz. 57, 236 (1994) [Phys. At. Nucl. 57, 222 (1994)]; L. A. Mikaelyan, Yad. Fiz. 65, 1206 (2002) 
[Phys. At. Nucl. 65, 1173 (2002)]. \\
4. H. Bethe, Proc. Cambridge Philos. Soc. 35, 108 (1935); G. V. DomogatskiT and D. K. Nadezhin, Yad. Fiz. 12, 1233 
(1970) [Sov J. Nucl. Phys. 12, 678 (1971)]. \\
5. M. B. Voloshin, M. I. VysotskiT, and L. B. Okun', Zh. Eksp. Teor. Fiz. 91, 754 (1986) [Sov. Phys. JETP 64, 446(1986)]. \\
6. S.A. Fayans, V. Yu. Dobretsov, and A. B. Dobrotsvetov, Phys. Lett. B 291, 1 (1992); V. Yu. Dobretsov, 
A. B. Dobrotsvetov, and S. A. Fayans, Yad. Fiz. 55, 2126 (1992) [SovJ.Nucl.Phys.55, 118 (1992)].\\
7. B. S. Dzhelepov {\it et al.} Beta Processes (Nauka, Leningrad, 1972). \\
8. S.A. Fayans, L. A. Mikaelyan, and V. V. Sinev, Yad. Fiz. 64, 1551 (2001) [Phys. At. Nucl. 64, 1475 (2001)]; 
V. I. Kopeikin et al., Yad. Fiz. 60, 2032 (1997) [Phys. At.Nucl.60, 1859(1997)].

\end{document}